# Detection Efficiency of a Spiral-Nanowire Superconducting Single-Photon Detector

D. Henrich, L. Rehm S. Dörner, M. Hofherr, K. Il'in, A. Semenov, and M. Siegel

*Abstract*— We investigate the detection efficiency of a spiral layout of a Superconducting Nanowire Single-Photon Detector (SNSPD). The design is less susceptible to the critical current reduction in sharp turns of the nanowire than the conventional meander design. Detector samples with different nanowire width from 300 to 100 nm are patterned from a 4 nm thick NbN film deposited on sapphire substrates. The critical current $I_C$ at 4.2 K for spiral, meander, and simple bridge structures is measured and compared. On the 100 nm wide samples, the detection efficiency is measured in the wavelength range 400-1700 nm and the cut-off wavelength of the hot-spot plateau is determined. In the optical range, the spiral detector reaches a detection efficiency of 27.6%, which is ~1.5 times the value of the meander. In the infrared range the detection efficiency is more than doubled.

*Index Terms*— Critical current, Nanoscale devices, Superconducting photodetectors.

## I. INTRODUCTION

SINCE the first demonstration of the detection principle [1], the Superconducting Nanowire Single-Photon Detector (SNSPD) has developed over the past decade into a mature technology [2]. Especially the prospect of detecting single photons in the infrared range with high efficiency is appealing for a wide variety of applications. The performance of SNSPD in this wavelength regime was improved in the past by a variety of different approaches, including improved coupling and absorption techniques, reduction of the nanowire cross-section and consideration of other superconducting materials. It was also found that the critical current of the devices plays an important role in the detection efficiency. The spectral dependence becomes broader if the bias current, which is limited by the critical current, can be adjusted closer to the de-pairing critical current limit [3].

Recently it was realized that the critical current is reduced by the geometrical layout of the detectors. It was observed, that with increasing fill factor the critical current of the devices is reduced [4]. This is due to the detector layout, where the nanowire is wound up like a meander with sharp turns at the end of each line to connect to the next line. The increase of the fill factor is realized by reduction of the gap width between the lines which leads to sharper turns. At the inner edge of such turns, the current density is increased, which lowers the energy barrier for vortex entry at this point. As soon as vortices overcome this barrier, they are immediately accelerated by the Lorentz force of the bias current which causes a voltage drop across the nanowire. Calculations of this current crowding effect based on the London theory were conducted by [5] for the case of thin film structures (film thickness $d \ll \xi$, $\xi$ coherence length) with widths $w \ll \lambda_{\text{eff}} = \lambda^2/d$, where $\lambda$ is the magnetic penetration depth. They allow a quantitative prediction of the reduction factor $R$ of the critical current with respect to the value for a straight line for simple geometries with only one turn.

The reduction was experimentally verified by measurements on NbTiN [6] and NbN [7] nanowires, but found to be weaker than theoretically expected. Additionally, the current trend to reduce the nanowire width of SNSPD further to reach higher detection efficiencies [8] should lead to less influence of the effect, since $R$ is expected to scale with $\xi/w$.

This paper aims to investigate the question, whether an improved detector layout with minimized critical current reduction due to sharp bends can improve the performance of the device. To this end, we have fabricated spiral detectors, as was suggested by [4], as well as conventional meander structures on NbN thin films and compare their critical current densities as well as the spectral detection efficiency.

## II. TECHNOLOGY

To directly compare different detector layouts without any contributions from material variation, samples are patterned on the same NbN thin film. The 4 nm thick films are deposited by dc reactive magnetron sputtering of a pure Nb target in a mixed Ar and $N_2$ atmosphere with partial pressure $p_{N2} = 2.9\ 10^{-4}$ mbar. The polished sapphire substrates are placed loosely on a heater plate that was kept at 800°C during deposition.

In a first lithography step, the detector layout was defined by electron beam exposure of a 90 nm thick resist film and etched by reactive ion etching in a $SF_6$+Ar plasma. In a second photo-lithography step, a 50 Ω impedance matched contact layout was added to provide connection to the rf readout.

Manuscript received October 10, 2012. This work was supported in part by the DFG Center for Functional Nanostructures under sub-project A4.3.

D. Henrich, L. Rehm, S. Dörner, M. Hofherr, K. Il'in and M. Siegel are with Institute of Micro- and Nanoscaled Systems, Karlsruhe Intitute of Technology, Karlsruhe, Germany (phone: +49 721 608 44994; fax: +49 721 757925; e-mail: dagmar.rall@kit.edu).

A. Semenov is with the Institute for Planetary Research, DLR, Berlin, Germany (e-mail: alexei.semenov@dlr.de).

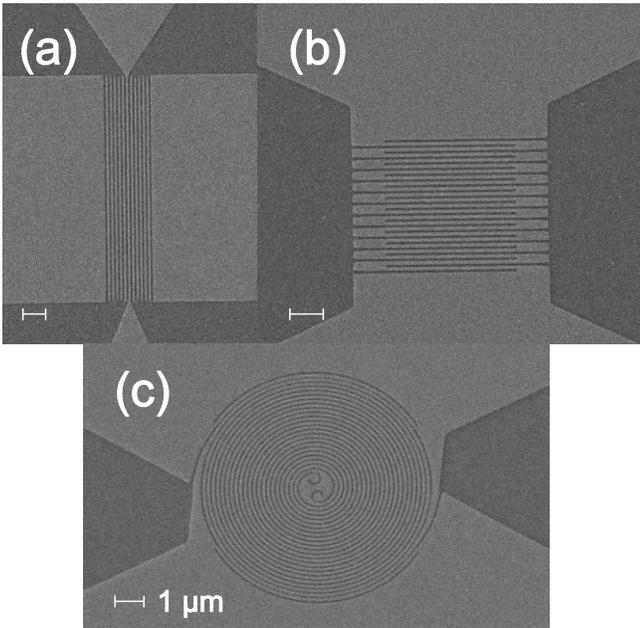

Fig. 1. SEM images of one sample set consisting of a simple bridge (a), a meander structure (b) and a spiral structure (c). Light parts are NbN, dark parts the substrate surface. The scale in each image marks 1 μm.

Fig. 1 shows scanning electron microscopy (SEM) images of a sample set consisting of a single bridge as a reference, a classical meander and a spiral detector. All samples in one set are structured with the same nanowire linewidth $w$ and gap width $g$ between the lines. The additional gap lines next to the bridge and at the top and bottom of the spiral are included to increase the uniformity of the linewidths and serve no further purpose. The spiral layout consists of two interlaced Archimedean spirals with inner radius $r_1$ and outer radius $r_2$. To exclude any current reduction at the ends of the spiral arms, the connections to the contact pads are structured with continuously increasing width. In a similar way, the inner ends are tapered and lead to the central area, where the effective width is larger than $w$. This is important to ensure that the central part does not limit the critical current density of the structure, since here a sharper turn is needed to connect the two spiral arms. As a drawback, this means that the central spot is not sensitive to incoming photons, which is something that would have to be improved for application purposes.

Three sets are made with a $w/g$ combination of 300 nm/100 nm, 200 nm/100 nm and 100 nm/60 nm. The first two sets are designed to study the influence of $w$, whereas the last set reflects typical values for actual detector devices and is used for the measurement of the spectral detection efficiency.

III. RESULTS

The samples were characterized by resistive four-point measurement from room temperature down to 4.2 K. The critical temperature $T_C$ was defined at the lowest temperature for which a non-zero resistance could be measured and was in average 10.8 K with only slight variation between the samples. No systematic dependence of $T_C$ on the sample layout was observed.

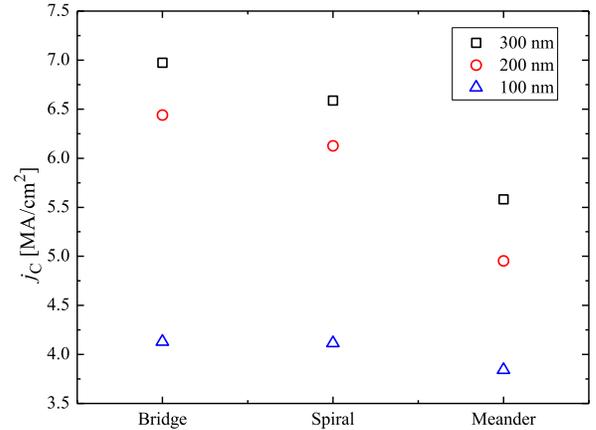

Fig. 2. Average Critical current densities $j_C$ measured at 4.2 K on bridge, spiral and meander structures for three different nanowire widths of 300 to 100 nm.

A. Critical current of spirals vs. meanders

The current-voltage characteristics of the samples are measured at 4.2 K by slowly ramping up the current. In all cases a well pronounced jump in the voltage defined the critical current of the sample. Fig. 2 shows the average critical current density for the three sample sets described in the last section. The highest $j_C$ value of 6.97 MA/cm$^2$ is measured on the bridges with $w = 300$ nm. The meander with the same widths show a considerably reduced value of 5.58 MA/cm$^2$, whereas the spirals have 6.59 MA/cm$^2$. Although this is also a lower value than for the bridges, it is an improvement of ~20% compared to the meander. The same trend is seen for the other sets but with overall lower values. The general reduction of $j_C$ for smaller structure widths is typical for very narrow nanowires (compare e.g. [9]).

In case of the 100 nm/60 nm set the meander have $j_C = 3.84$ MA/cm$^2$, which is also reduced with respect to the bridge $j_C = 4.13$ MA/cm$^2$, but the relative reduction in this case is lower, although the smaller $g$ should even lead to an increased effect [5]. However, this is in accordance with the prediction that the critical current reduction factor scales with $\xi/w$. The coherence length can be estimated from magnetic field measurements on unstructured NbN films, which for comparable thickness and deposition conditions gives values of 4.4 nm [3]. For a smaller $w$, $R$ is closer to unity and the critical current less reduced. Since the initial reduction is weaker, the relative improvement by the spiral is also lower. With a value of 4.11 MA/cm$^2$, which is almost the same as in the bridge, the spiral structure has a 7% increased $j_C$ compared to the meander. When considering even smaller wire widths, we expect that the difference in $j_C$ between a spiral and a meander structure will soon become smaller than the variation due to fabrication.

B. Detection efficiency

The spectral detection efficiency is measured on a spiral and a meander from the 100 nm/60 nm set at identical measurement conditions. For the other sets, the cross-section is too large so that the hot-spot plateau is shifted into the UV range [10]. The wavelength of the incident light can be changed continuously by a monochromator in the range 400-1700 nm. After the filter stages the light is coupled into a multimode optical fiber to feed it into the cryostat. The fiber ends at a position above the sample that can be adjusted in the cooled down state. The sample is fixed with conductive silver paste on a copper holder which is thermally coupled to the liquid Helium bath. The electrical connections are wire bonded to an rf-circuit board on the sample holder, where the signal is lead through a bias tee and noise filters to a semirigid rf cable. At room temperature, the signal is amplified by 70 dB with an effective bandwidth of 1.7 GHz and then sent to the pulse counter. The bias current is supplied by a low-noise battery source.

Before each measurements, it is verified that the incident light does not influence on the critical current of the device, i.e. does not introduce a significant local temperature change, and that the number of counts registered by the detector scales linearly with the light intensity. The intensity is then fixed at a value which corresponds to ~5 pW on the detector area, exact number varying with wavelength. The detector area $A_\text{det}$ is defined for the meander sample as the rectangular area 4.2 μm × 4.2 μm around the meander. For the spiral it is the circle around the structure with exclusion of the inner blind spot $A_\text{det} = \pi(r_2^2 - r_1^2)$ with $r_2 = 4.2$ μm and $r_1 = 0.6$ μm. Including the inner part of the spiral into the definition only changes the $DE$ values by less than a percentage point. The rate of photons $n_\text{ph}$ passing $A_\text{det}$ in each case and for each wavelength is determined by careful calibration. The dark count rate $DCR$ of the devices is determined by blocking the beam path and shows equal dependence on the bias current $I_\text{B}$ for both devices. At $I_\text{B} = 0.95 \cdot I_\text{C}$, the $DCR$ is ~1500 s$^{-1}$.

For each given wavelength λ and bias current $I_\text{B}$, the rate of counts $n$ is measured and the detection efficiency is then defined as

$$DE = \frac{n - DCR(I_B)}{n_{ph}}. \quad (1)$$

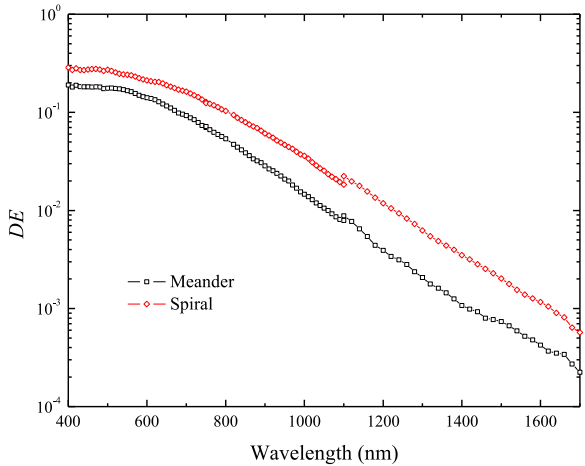

Fig. 3. Spectral detection efficiency of a meander (black) and a spiral (red) nanowire detector measured at $I_\text{B} = 0.95 \cdot I_\text{C}$ and $T = 4.2$ K. The offset at λ = 1100 nm is caused by a filter change.

TABLE I
CUT-OFF WAVELENGTH OF SPIRAL AND MEANDER DETECTORS

| $I_\text{B}/I_\text{C}$ | Meander | Spiral |
|---|---|---|
| 0.95 | 620.0 | 632.8 |
| 0.90 | 554.3 | 576.6 |
| 0.85 | 494.0 | 525.3 |
| 0.80 | 455.8 | 477.2 |

Cut-off wavelengths $\lambda_\text{C}$ in nm determined by fitting of (3) to the normalized spectral detection efficiency.

The spectral dependence of $DE$ for $I_\text{B} = 0.95 \cdot I_\text{C}$ and $T = 4.2$ K is shown in Fig. 3 for a meander and a spiral detector in direct comparison. In the whole spectral region of measurements, the spiral shows a higher $DE$ than the meander. In the hot-spot plateau region between roughly 400 and 600 nm, the meander reaches a detection efficiency of 18.2%, whereas the spiral design achieves 27.6%. In the infrared range, which is especially interesting for telecom applications, the $DE$ is more than doubled: For 1540 nm, $DE = 0.06$% for the meander and $DE = 0.16$% for the spiral.

## IV. DISCUSSION

The absolute $DE$ for a fixed wavelength depends on the ratio of $I_\text{B}/I_\text{C}$, which is a common observation in experimental investigations of SNSPD. It first rises fast with $I_\text{B}/I_\text{C}$ and then saturates at a certain bias current, which depends on the actual wavelength. For the meander, $I_\text{C}$ is limited by the sharp turns at the edges of the detector area. Here, the condition of $I_\text{B}/I_\text{C} = 0.95$ is met only locally in the edges, whereas for the straight portions of the nanowires, $I_\text{B}/I_\text{C}$ can be much lower. This means that the $DE$ on the main, central part of the detector is effectively reduced as compared to the areas near the turns. The spiral design on the other hand also results in an $I_\text{C}$ value less than in a straight nanowire, but larger than $I_\text{C}$ of the meander. Furthermore, this reduction is the same along the whole spiral line. Thus, the higher local detection efficiency is reached on a large part of the detector area, leading to higher overall $DE$ values.

The spectral dependences of the detection efficiency of both samples look quite similar. They have flat regions up to ~600 nm and fall steadily at larger wavelengths. According to the hot-spot model [11], the cut-off wavelength $\lambda_\text{C}$ describing this transition depends on the bias current:

$$\lambda_C \propto wd\left(1 - \frac{I_B}{I_C^d}\right) \quad (2)$$

Here, $I_\text{C}^d$ is the de-pairing critical current, which marks the material limit of the super-current. Since both devices are structured from the same film, $I_\text{C}^d$ is identical. However, the experimental critical current $I_\text{C}$ of the device is reduced below this value by the current crowding effect and is lower for the meander. Consequently, if both devices are biased at $0.95 \cdot I_\text{C}$, the effective ratio $I_\text{B}/I_\text{C}^d$ is smaller for the meander and a reduced value of $\lambda_\text{C}$ is expected. To eliminate the influence of the wavelength-dependent absorptance on the $DE$, the spectra are measured at different bias currents and then normalized to the spectra for $I_\text{B} = 0.95 \cdot I_\text{C}$ (see Fig. 4). From these normalized spectra, the $\lambda_\text{C}$ of the intrinsic detection efficiency can be evaluated by a least mean square fit of

$$\frac{DE(I_{B,1})}{DE(I_{B,2})} = \frac{1+(\lambda/\lambda_{C,2})^{p_2}}{1+(\lambda/\lambda_{C,1})^{p_1}} \quad (3)$$

to the experimental data, where $\lambda_{C,x}$ and $p_x$ are fitting parameters for the bias current $I_{B,x}$ (for details see [3]). The respective fits are shown for the spiral detector as solid lines in Fig. 4. The values of $\lambda_C$ determined in this way for four bias currents on each sample are shown in Table I. As expected from the hot spot model, the values are larger for the spiral structure. However, since the difference in $I_C$ between the structures was small, the shift in the cut-off wavelength is also only minor.

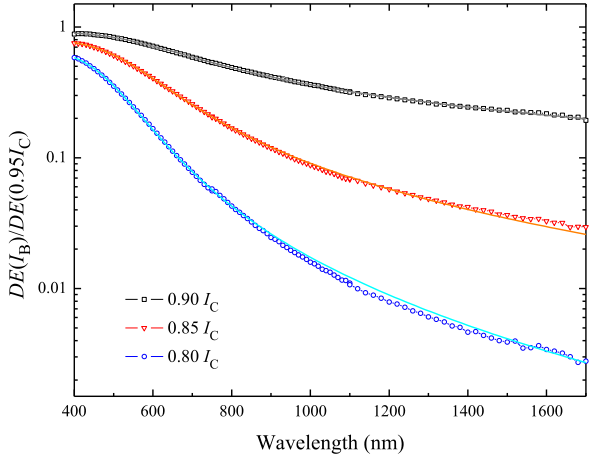

Fig. 4. Spectral detection efficiency of a spiral detector with $w = 100$ nm, normalized point-by-point to the detection efficiency at $I_B = 0.95 \cdot I_C$. The solid lines are fits of (3) to the data.

Furthermore, we note that the normalized detection efficiency shown in Fig. 4 does not run to 1 in the hot spot regime, meaning that for the given measurement conditions, $DE(I_B/I_C)$ is not yet in saturation. The normalized spectral $DE$ of the meander sample, also measured for 4.2 K, showed very similar dependence. The same argument for the reduced overall $DE$ as given above thus applies also to the hot-spot regime.

An added benefit of the spiral detector structure is that its detection efficiency is almost independent on the polarization of incident light. For the classical meander design, the absorption probability depends on the polarization relative to the parallel nanowires [12,13], because the wires act as a wire-grid polarizer. The polarization-dependence of spiral detectors has been experimentally investigated by [14]. The detection efficiency to polarized 650 nm irradiation was measured on NbTiN SNSPDs with spiral and meander layout. It was shown that the degree of polarization is reduced significantly for the spiral layout. In both cases, the detection efficiency with respect to randomly polarized light is the mean value of the parallel and perpendicular detection efficiencies. For our measurements, where the degree of polarization of the light is below 2%, we thus expect no difference in $DE$ due to the polarization. Indeed in [14], the measurements reveal that for the randomly polarized case the detection efficiency of the spiral is even lower than for the meander. The reason for this is that in the design of the spiral, the turn in the center, where the inwards spiral changes to the second spiral going out, is once again structured as a sharp turn as in the meander design (compare Fig. 1c) here with Fig. 1b) of [14]). The critical current of the structure is limited by this turn, leaving the whole rest of the spiral with a lower $I_B/I_C$ ratio and consequently a lower detection efficiency. This consideration agrees well with our measurements and further supports the assumption, that in the meander design the photons are mainly detected by the straight part of the detector due to the much larger area as compared to the edges.

## V. Conclusions

A spiral SNSPD detector layout with improved uniformity of the current distribution was developed and fabricated from a NbN thin film. Measurements of the critical current showed that the new design suffers less from the reduction of $I_C$ by the current crowding effect. For a line-width of 300 nm an increase of ~20% in the critical current with respect to a conventional meander structure was achieved. However, as the width of the nanowire is reduced, the effect becomes weaker in accordance with theoretical models.

For structures with $w = 100$ nm and $g = 60$ nm, the spectral detection efficiency was measured in the wavelength range 400-1700 nm. Since the spiral shows only a slightly larger $I_C$, the cut-off wavelength is not significantly larger. However, the more uniform current distribution leads to an increase of the detection efficiency, especially so at large wavelengths. For optical photons, the detection efficiency was improved by a factor of 1.5 from 18.2% to 27.6%. In the infrared range, at 1540 nm an even larger increase in the detection efficiency by a factor of 2.7 was achieved. The additional benefit of a reduced dependence to the polarization and more convenient optical coupling to the incoming light due to the round shape should make such spiral SNSPD layouts attractive for many applications.